\documentclass[a4paper,11pt]{article}
\usepackage{pos}
\usepackage[T1]{fontenc} % if needed
\usepackage{subfig}
\usepackage{aas_macros} % needed to resolve ADS journal abbreviations
\usepackage{tabularx}  % needed for tables
\usepackage{longtable}  % 
\usepackage{booktabs} % 
\usepackage{units} %
\usepackage{comment}
\usepackage{xspace}
\usepackage{lineno}
\usepackage{multirow}
\usepackage{siunitx}

\newcommand{\gammapy}{\textsc{gammapy}\xspace}
\newcommand{\ctools}{\textsc{ctools}\xspace}

\def\LIV{\ifmmode {\mathrm{LIV}}\else{\scshape LIV}\fi\xspace}
\def\LI{\ifmmode {\mathrm{LI}}\else{\scshape LI}\fi\xspace}

\title{CTA sensitivity for probing cosmology and fundamental physics with gamma rays}
 \ShortTitle{CTA sensitivity for probing cosmology and fundamental physics}
 
\author*[a]{I. Vovk}
\author[b]{J. Biteau}
\author[c]{H. Martinez-Huerta}
\author[d]{M. Meyer}
\author[e]{S. Pita}
\affiliation[a]{Institute for Cosmic Ray Research, The University of Tokyo\\
5-1-5 Kashiwa-no-Ha, Kashiwa City, Chiba, 277-8582, Japan}
\affiliation[b]{Laboratoire de Physique des 2 Infinis Ir\`ene Joliot Curie, \\
CNRS/IN2P3, Universit\'e Paris-Sud, Universit{\'e} Paris-Saclay, Orsay, France}
\affiliation[c]{Department of Physics and Mathematics, Universidad de Monterrey,\\
Av. Morones Prieto 4500, 66238, San Pedro Garza Garc\'ia NL, M\'exico}
\affiliation[d]{University of Hamburg, Institute for Experimental Physics, \\
Luruper Chaussee 149, D-22761 Hamburg, Germany}
\affiliation[e]{APC, Universit\'e Paris Diderot, \\
CNRS/IN2P3, CEA/IRFU, Obs. de Paris, Sorbonne Paris Cit\'e, France }

\forColl{CTA Consortium} % W/O "Collaboration"

\emailAdd{vovk@icrr.u-tokyo.ac.jp}
\emailAdd{jbiteau.pro@gmail.com}
\emailAdd{humberto.martinezhuerta@udem.edu}
\emailAdd{mmanuel.e.meyer@fau.de}
\emailAdd{pita@apc.in2p3.fr}
\abstract{
The Cherenkov Telescopic Array (CTA), the next-generation ground-based gamma-ray observatory, will have unprecedented sensitivity, providing answers to open questions in gamma-ray cosmology and fundamental physics. Using simulations of active galactic nuclei observations foreseen in the CTA Key Science Program, we find that CTA will measure gamma-ray absorption on the extragalactic background light with a statistical error below 15\% up to the redshift of 2 and detect or establish limits on gamma halos induced by the intergalactic magnetic field of at least 0.3~pG. Extragalactic observations using CTA also demonstrate the potential for testing physics beyond the Standard Model. The best state-of-the-art constraints on the Lorentz invariance violation from astronomical gamma-ray observations will be improved at least two- to threefold. CTA will also probe the parameter space where axion-like particles can represent a significant proportion -- if not all -- of dark matter. Joint multiwavelength and multimessenger observations, carried out together with other future observatories, will further foster the growth of gamma-ray cosmology.
}

\FullConference{37$^{\rm{th}}$ International Cosmic Ray Conference (ICRC 2021)\\
		July 12th -- 23rd, 2021\\
		Online -- Berlin, Germany}

\begin{document}
\maketitle

\section{Introduction}

Study of $\gamma$-ray propagation from bright and distant astrophysical emitters at very-high energies (VHE, $E > 30\,$GeV) over cosmological distances has emerged as a successful branch of ground-based gamma-ray astronomy over the past decade. Gamma~rays emitted by extragalactic sources (e.g. blazars) can interact along their way to the observer, producing of $e^+e^-$ pairs on near-UV to far-infrared photon fields.
% \cite{REF::NIKISHOV::JETP1962,1967PhRv..155.1404G, 1967PhRv..155.1408G}. 
This results in an absorption horizon for $\gamma$ rays, beyond which the received emission is suppressed \cite[e.g.][]{2008A&A...487..837F}.
% \cite{2008A&A...487..837F, 2011MNRAS.410.2556D, 2012MNRAS.422.3189G}.
This effect makes it possible to probe the extragalactic background light (EBL), populating the voids of the Large Scale Structure. The uncertain specific intensity of EBL has been shown to agree with expectations from galaxy counts at the $\sim$\,30\,\% level (for EBL wavelengths up to a few tens of $\mu$m) based on the data from the current-generation $\gamma$-ray observatories (H.E.S.S., MAGIC, VERITAS)~(e.g. \cite{2021JCAP...02..048A} and references therein).
% \cite{2013A&A...550A...4H, 2015ApJ...812...60B, 2016A&A...590A..24A,2017A&A...606A..59H,2019ApJ...885..150A,2019ApJ...874L...7D}.
The redshift evolution of EBL however remains poorly constrained by the ground-based telescopes due to small number of $\gamma$-ray sources detected beyond $z\sim 0.5$.

Absorption of $\gamma$ rays on EBL results in generation of $e^+e^-$ pairs, whose propagation is subject to the intergalactic magnetic field (IGMF). The basic properties of IGMF -- strength and coherence length -- as well as its origin, remain poorly constrained 
\cite{2013A&ARv..21...62D}.
% \cite{2013A&ARv..21...62D, 2019Galax...7...47S}. 
These pairs reprocess their emission via Inverse Compton scattering of the cosmic microwave background (CMB) photons, resulting in a lower-energy $\gamma$-ray signal. The non-detection of the latter in spectral and spatial searches by current $\gamma$-ray telescopes has constrained the IGMF strength to be ${\gtrsim}\,10^{-14}$~Gauss for a coherence length larger than a megaparsec and blazar duty cycles $\gtrsim 10^5$\,years; see \cite{2021JCAP...02..048A} and references therein. The same $e^+e^-$ pairs could also lose their energy through plasma instabilities, whose relative strength to Inverse Compton cooling is under active theoretical debate.

Propogation of $\gamma$-rays could also be altered in non-standard scenarios, such as Lorentz invariance violation (LIV) 
% \cite{1999ApJ...518L..21K, 2008PhRvD..78l4010J}
or coupling of $\gamma$ rays to axion-like particles (ALPs) inside magnetic fields.
% (e.g., \cite{deangelis2007,mirizzi2007,deangelis2011}). 
These scenarios may result in reduction of the opacity of the Universe to multi-TeV gamma ray propagation 
% \cite{1999ApJ...518L..21K, 2008PhRvD..78l4010J} 
and specific spectral signatures of Active Galactic Nuclei (AGN) embedded in the magnetic field of their parent clusters;
% \cite{hess2013:alps, ajello2016,zhang2018}
see \cite{2021JCAP...02..048A} for a deeper review.
% The former effect has been already constrained with the data of existing ground-based telescopes \cite{2015ApJ...812...60B,2019ApJ...870...93A}.

In the near future, Cherenkov Telescope Array (CTA)
% , owing to its unprecedented sensitivity, angular and energy resolutions,
will open a new page in the studies of VHE blazars. In what follows, we investigate how the CTA AGN \cite{2019scta.book..231Z} and Cluster of Galaxies \cite{2019scta.book..273Z} Key Science Projects can be used to probe $\gamma$-ray cosmology. We explore the CTA potential in the so-called Alpha and Omega configurations to measure the EBL imprint in the blazar spectra and to constrain or detect IGMF, ALPs and LIV signatures with deep targeted observations of distant AGNs.

\section{CTA measurement of EBL intensity}\label{sec:source_and_ebl}

% Interactions of $\gamma$ rays with \textbf{the EBL} result in an effective \textbf{opacity} of the Universe that depends on the distance of the $\gamma$-ray source and on the $\gamma$-ray energy. The \textbf{interaction} process involved is pair production of electrons and positrons.
% The EBL photon density, $\partial n / \partial \epsilon$, can be parametrized to estimate the  optical depth to $\gamma$ rays, $\tau(E_\gamma,z_0)$ from $\gamma$-ray data. Best-fit EBL parameters are obtained marginalizing over the intrinsic source spectral parameters when modeling the $\gamma$-ray source spectrum.
 
The specific intensity of EBL can be parametrised by its optical depth $\tau(E_\gamma)$ to the penetrating $\gamma$ rays. Constraints on it have already been derived with current $\gamma$-ray telescopes assuming its linear scaling with respect to a given EBL-model, i.e. $\tau'(E_\gamma) = \alpha\times\tau(E_\gamma)$
% (e.g., \cite{2013A&A...550A...4H,2015ApJ...815L..22A, 2016A&A...590A..24A, 2012Sci...338.1190A}).
(e.g. \cite{2012Sci...338.1190A}).
In this work, we adopt a similar parametrization in order to illustrate the overall performance of CTA compared to the currently existing instruments and assess its capability to constrain the redshift dependence of the normalization coefficient, $\alpha$. To this end, out of the database of VHE AGN spectra expected from the CTA AGN KSP, we select a list of candidates expected to be detected at energies beyond the cosmic $\gamma$-ray horizon. 
% that would best constrain $\gamma$-ray absorption on cosmic scales.
In total, we simulate around $830$~hours of CTA observations.

% \subsection{Source selection from AGN KSP}\label{subsect::ebl_ksp_sel}

% The CTA KSP dedicated to AGN involves three different observing programs: 
% \begin{itemize}
%     \item \textbf{long-term monitoring} of selected AGNs that will be observed weekly in 30 minute snapshots; a list of potential targets is presented in chapter 12 of \cite{2019scta.book.....C}.
%     
%     \item \textbf{high-quality spectra} program with deep observations of AGNs of different classes and at different distances. A list of possible targets is given in \cite{2019scta.book.....C}. In this work, we study here blazars of interest for EBL studies from the 3FHL catalogue \cite{2017ApJS..232...18A}.
%     
%     \item \textbf{blazar flares follow-up} program target VHE flares from AGNs, triggered either by external facilities or internally by the monitoring program performed with CTA.
% \end{itemize}

% Out of this rich database of VHE AGN spectra expected from the CTA KSP dedicated to AGNs~\cite{2019scta.book.....C}, we select a list of candidates that best constrain $\gamma$-ray absorption on cosmic scales.
% In total, we simulate $830\,$hours of CTA observations of \textbf{AGNs} expected to be detected at energies beyond the cosmic $\gamma$-ray horizon with a class-dependent cut-off. This observation budget represents about a quarter of the AGN KSP. Cumulative constraints on $\gamma$-ray propagation could also be expected from the remaining three quarters \textbf{of the total live-time}, although likely in a subdominant manner.

% \subsection{Determination of the optical depth}

In order to reconstruct the scale factor $\alpha$ of the benchmark EBL model, we simulate both signal and background counts for each of the blazars selected above and marginalize over their respective intrinsic spectral parameters.
\begin{figure}[t]
    \center
    \includegraphics[width=0.8\columnwidth]{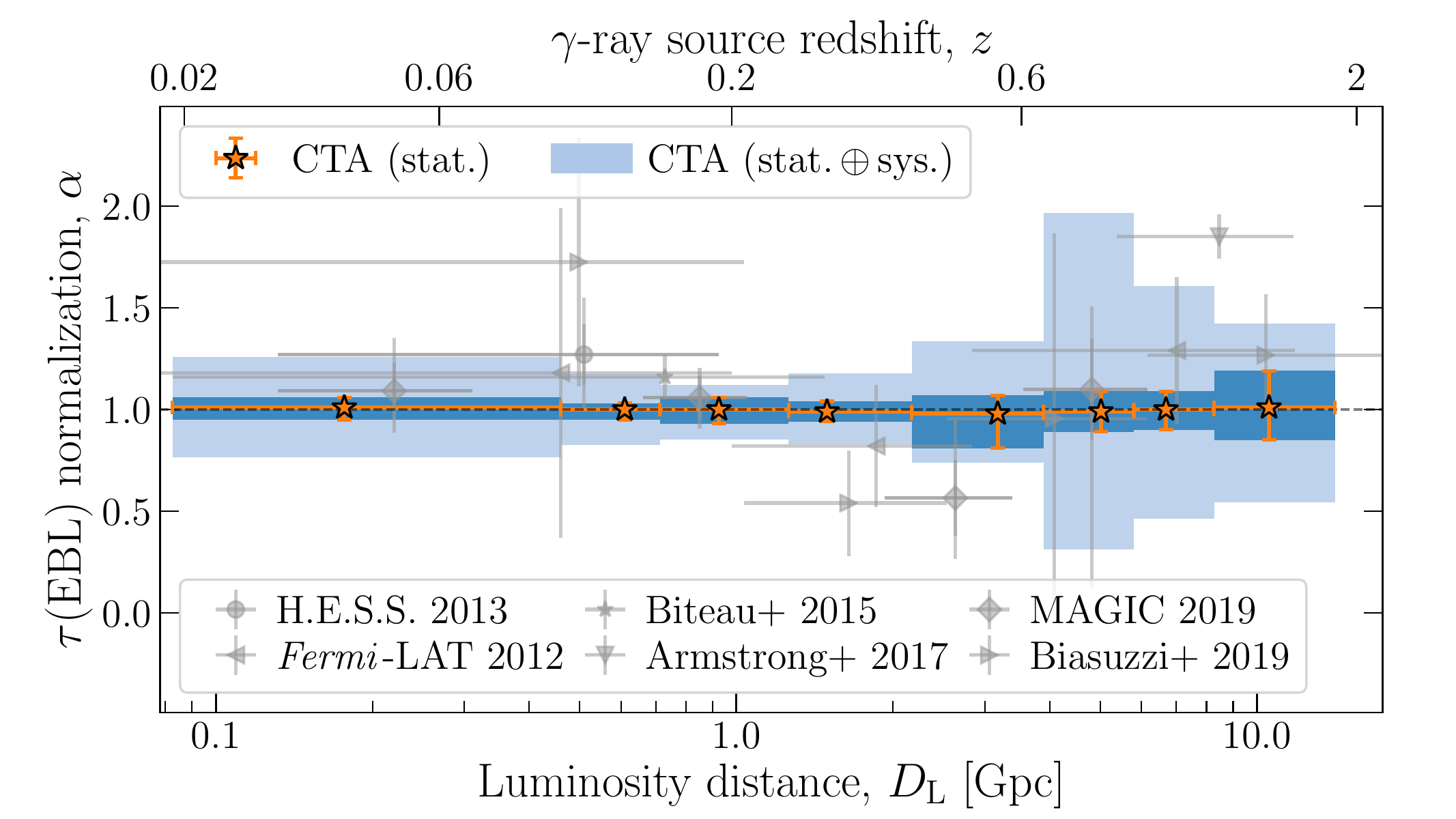}
    \caption{
        Reconstructed EBL scale factor as a function of redshift.
%         The 16\% and 84\% quantiles of the derived scale $\alpha$ distribution are depicted with orange error bars and dark blue shaded bars; note that these are larger than the uncertainty on the distribution's mean value.
%         The total -- statistical plus systematic -- uncertainty is shown in light blue. 
%         Shown constraints from the literature are extracted from Ref.~\cite[][]{2019A&A...627A.110B} and are grouped by the instrument in the figure legend (ground-based instruments in the top row, \textit{Fermi}-LAT -- in the bottom). 
        Extracted from~\cite{2021JCAP...02..048A}.
    }
    \label{fig::ebl_constrains1}
\end{figure}
To estimate the uncertainties on $\alpha$, we generate 1000 realizations of the source spectra in each redshift bin and reconstruct the optical-depth normalization for each of these. The systematic uncertainties are estimated via a similar procedure employing the Instrument Response Functions (IRF) bracketing approach.

The obtained results are summarized in Fig.~\ref{fig::ebl_constrains1}. The optical depth scale factor $\alpha$ is reconstructed with an average statistical uncertainty of $<15\%$. The systematic uncertainty, however, varies from $12\%$ to $50\%$ depending on the redshift. One may note that uncertainties on $\alpha$ resulting from IRF bracketing up to $z=0.65$ are comparable to those stemming from the state-of-the-art EBL models.
% \citep[{e.g.},][]{2011MNRAS.410.2556D,2010ApJ...712..238F}.
% \citep[e.g.][]{2011MNRAS.410.2556D}.

In spite of its simplicity, the performed analysis provides a first illustration of what CTA is expected to deliver. Still, at small redshifts, constraints from blazars with cut-offs at 10\,TeV will crucially affect the CTA capability to probe the cosmic infrared background component up to 100$\,\mu$m, a wavelenth range that is still under-constrained. Low-energy observations of $\gamma$-ray sources beyond redshift $z\sim 0.5$ will be important to constrain interactions with UV photons down to 0.1$\,\mu$m. The CTA low-energy capabilities will also be crucial to constrain the cosmic star formation history, particularly up to its peak located at $z\sim 1.5-2.5$.
% , already probed by \textit{Fermi}-LAT data \cite{2018Sci...362.1031F}. 
High-precision CTA measurements combined with a large source sample detected beyond $z=1$ with \textit{Fermi}-LAT may make it possible to probe not only the EBL spectrum at $z=0$ in the wide $0.1-100\,\mu$m range, but also its evolution over cosmic time, including -- by means of the integral nature of EBL -- contributions from distant UV sources beyond $z \sim 2$.

Though AGN observations will be essential to constrain the EBL spectrum and evolution, complementary constraints on cosmological parameters, such as $H_0$ and $\Omega_M$, can be expected as well. 
Indeed, the optical depth $\tau(E_\gamma)$ is roughly proportional to the EBL density and inversely proportional to $H_0$ (e.g. \cite{1994ApJ...423L...1S}).
% \cite{1994ApJ...423L...1S, 2005APh....23..598B, 2013ApJ...771L..34D, 2015ApJ...812...60B, 2019arXiv190312097D}.
Consequently, $H_0$ uncertainty at least as large as that on the scale factor $\alpha$ may be expected for an EBL spectrum fixed to the level expected from galaxy counts. Dedicated studies will be required to assess the full potential of CTA to constrain cosmological parameters.

\section{CTA sensitivity to IGMF}\label{sect::IGMF}

 CTA observations promise to address at once several aspects of IGMF influence on the VHE appearance of blazars: time delay of the cascade emission,
%  ~\citep{2004ApJ...613.1072R, Murase08, Ichiki08, NeronovSemikoz09}
 presence of broad spectral features due to the cascade contribution,
%  ~\citep{Neronov10, Taylor11, Tavecchio11}
 and extended emission around otherwise point-like source;
%  ~\citep{HEGRA_IGMF, MAGIC_IGMF, 2014A&A...562A.145H, 2017ApJ...835..288A}.
 see \cite{2021JCAP...02..048A} for a deeper review.

Here, we restrict ourselves to IGMF strengths which could result in spectral and morphological signatures in $\gamma$-ray observations, i.e.\ higher than those probed with time delays. We perform simulations of the prototypical extreme blazar 1ES\,0229+200, which, owing to its hard intrinsic $\gamma$-ray spectrum extending to $\sim 10\,$TeV, and the lack of strong $\gamma$-ray variability, is perhaps one of the best-suited sources for cascade signatures searches.

We use the \textsc{CRPropa} code\footnote{\url{https://crpropa.desy.de/}}
% ~\citep{CRPropa} 
to simulate the development of electromagnetic cascades, assuming a randomly oriented IGMF in uniform cells of 1\,Mpc size with the field strength fixed in each simulation. We employ the simplifying assumption, for 1ES\,0229+200, of a conical jet with a $10^\circ$ opening angle, tilted by an angle of $5^\circ$. The source intrinsic spectrum is taken to be a power law with an exponential cut-off at $E_\mathrm{cut}' = 10$\,TeV.
% , in line with the lower limit set by current-generation VHE observations~\citep{HESS_1ES0229, VERITAS_1ES0229}.
Finally, we use \ctools to simulate a 50-hr long CTA observation and to compute the likelihood for a given set of spectral parameters for each tested IGMF setup.

% \subsection{CTA sensitivity to IGMF}

The CTA sensitivity to IGMF-induced effects is quantified combining both spectral and morphological information. This combination is enabled by updating the extended emission model at each step of the fit according to the point source spectral parameters, using a pre-computed cascade emission library for various IGMF strengths. The detection significance of the cascade component is thus computed self-consistently accounting for both its spectral and morphological features.

\begin{figure}[t]
    \center
    \includegraphics[width=0.73\columnwidth]{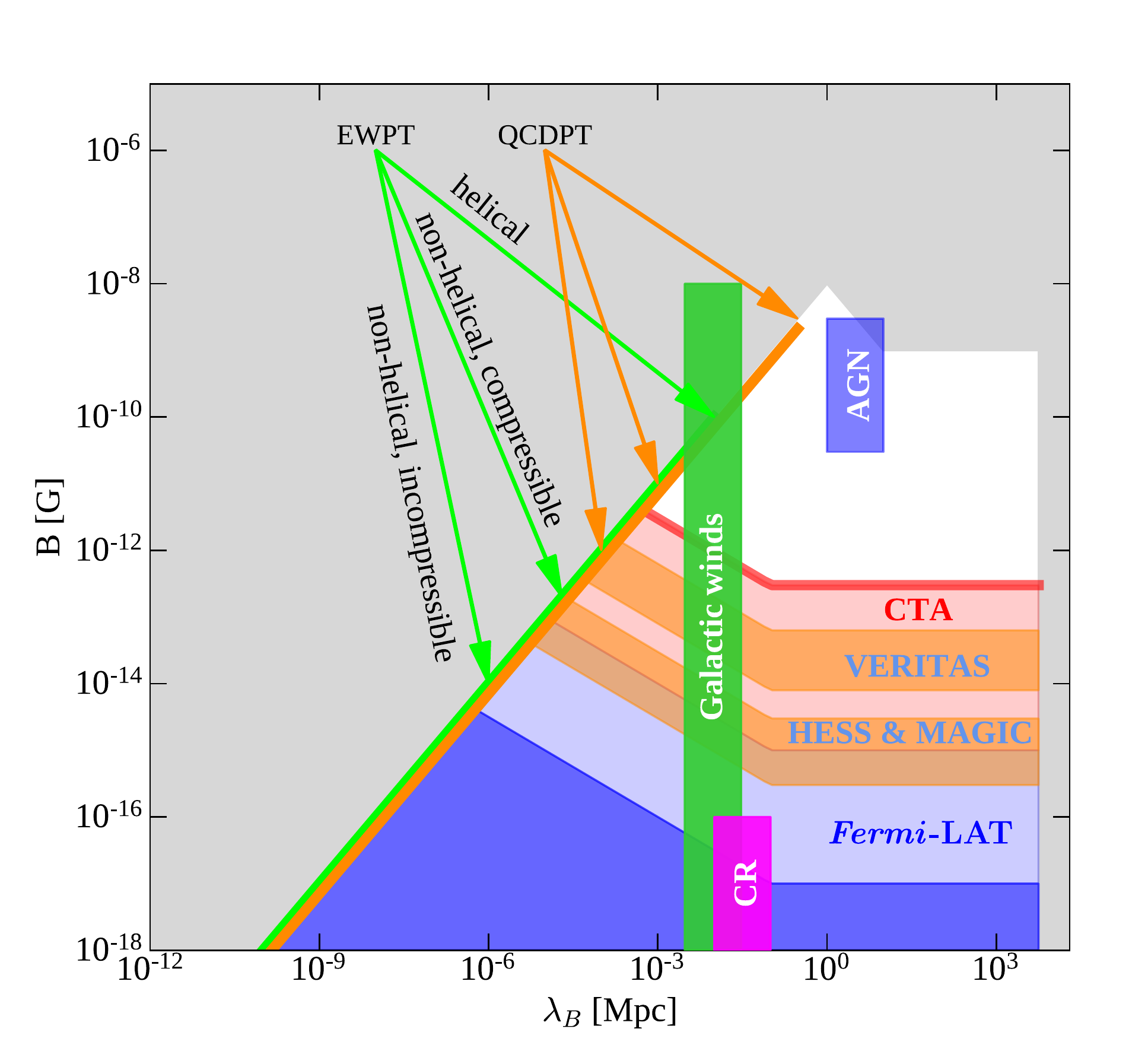}
    \caption{
        Sensitivity of CTA to IGMF signatures compared to existing observational constraints and theoretical predictions as a function of the IGMF strength $B$ and coherence scale $\lambda_{B}$. 
%         The figure is adapted from~\citep{2013A&ARv..21...62D}, with addition of the \emph{Fermi}-LAT excluded regions from~\citep{Taylor11} and~\cite{2018ApJS..237...32A} shaded with dark and light blue correspondingly; exclusion regions from H.E.S.S., MAGIC~\cite[overlapping,][]{2014A&A...562A.145H, 2010A&A...524A..77A} and VERITAS~\cite{2017ApJ...835..288A} are shown in orange. 
        The red line marks the maximal IGMF strength that would be detectable at $\gtrsim 5\sigma$ level in a 50\,hour long CTA observation of 1ES\,0229+200, assuming the $\sim 10^7$~yr source activity.
%         with a $10^\circ$-wide jet inclined by $5^\circ$ with respect to the line of sight. 
%         The dashed blue line denotes the \emph{Fermi}-LAT lower limit on IGMF strength from Ref.~\citep{2018ApJS..237...32A}, taken within the validity range of their simulations and corresponding to a 10\,year source activity time. 
%         Grey regions are disfavored by direct probes and theoretical considerations. The orange and green lines represent theoretically favored regions for the generation of the IGMF during either electro-weak (EWPT) or QCD (QCDPT) phase transitions~\citep[][and references therein]{2013A&ARv..21...62D}. 
        The white region is beyond the sensitivity of the instruments discussed here. 
%         Filled vertical boxes show favored regions of models where the IGMF is generated by a frozen-in \emph{B}-field, originating from AGN outflows~\cite{Furlanetto:2001}, galactic winds~\citep{Bertone:2006} or induced by cosmic-ray streaming~\citep{Miniati:2011}.
        Extracted from~\cite{2021JCAP...02..048A}.
    }
    \label{fig::igfm_exclusion_plot}
\end{figure}

It follows that CTA will be able to detect a cascade emission, provided that the IGMF strength smaller than $2 \times 10^{-13}$\,G (for an aligned jet and a coherence length of 1\,Mpc). CTA measurements will thus almost close the gap between the existing IGMF constraints
% \citep{HEGRA_IGMF, MAGIC_IGMF, 2014A&A...562A.145H,2017ApJ...835..288A} 
and the maximal field strength consistent with galaxy formation models
% ~\cite{dolag05, donnert09}
~\cite{2021JCAP...02..048A}. The derived sensitivity region where the IGMF could be detected with CTA is shown in Fig.~\ref{fig::igfm_exclusion_plot} along with the existing constraints from various instruments.

It should be noted that blazar duty cycles shorter than $\sim 10^7$~yr would substantially reduce CTA sensitivity to IGMF-induced signatures -- e.g. a 30-fold IGMF limit degradation was found in \cite{2018ApJS..237...32A} for an activity time scale reduced from $10^7$ to $10^4$ and from $10^4$ to $10$ years. A similar degradation should also apply here. On the other hand, a combination of CTA and \emph{Fermi}-LAT data could further broaden the probed parameter space with suitable AGNs, providing contemporaneous observations that are required for variable $\gamma$-ray sources.

% In conclusion, our estimates suggest that among the existing and planned \textbf{IACTs,} CTA will provide the best sensitivity to field strengths up to $3\times 10^{-13}$\,G, probing a range where the multiple different mechanisms may be responsible for IGMF origin (\cite{2013A&ARv..21...62D} and references therein).

\section{CTA sensitivity to ALP signatures in NGC\,1275 observations}
\label{sec:axions}

% The propagation of $\gamma$ rays could be affected by interactions with yet undiscovered particles beyond the Standard Model of particle physics. In particular, $\gamma$ rays could oscillate into ALPs in ambient magnetic fields.
% Photon-ALP conversions can lead to distinctive signatures in AGN spectra. On the one hand, ALPs evade pair production with photon fields such as the EBL and the photon-ALP oscillation can thus significantly reduce the effective optical depth. Indications for such a reduction could have been found in blazar observations both with ground-based instruments and \emph{Fermi}-LAT~\cite{deangelis2007,deangelis2011,horns2012,essey2012,rubtsov2014,kohri2017} and were interpreted as evidence for ALPs~\cite{mirizzi2007,deangelis2007,sanchezconde2009,deangelis2011,dominguez2011alps,meyer2013}. However, recent analyses found that $\gamma$-ray spectra are in general compatible with predictions from attenuation on the EBL~\cite{sanchez2013,dominguez2015,2015ApJ...812...60B}. Such effects will be actively probed with CTA~\cite{2013JCAP...11..023M, meyer2014cta,wouters2014,2015PhLB..744..375T}.

Gamma ray to ALP conversions can lead to distinctive signatures in AGN spectra, including a reduced effective optical depth
% ~\cite{mirizzi2007,deangelis2007,sanchezconde2009,deangelis2011,dominguez2011alps,meyer2013} 
and oscillatory patterns in AGN spectra that depend on the morphology of the traversed magnetic fields.
% ~\cite[][]{ostman2005,wouters2012}.
% On the other hand, at energies around the critical one, marking the transition to the so-called strong mixing regime, oscillatory patterns are expected in AGN spectra that depend on the morphology of the traversed magnetic fields~\cite[][]{ostman2005,wouters2012}. 
The search for such features at $\gamma$-ray energies has already resulted in the strongest bounds on the photon-ALP coupling to date for the ALP masses between $4\,\mathrm{neV}$ and $100\,\mathrm{neV}$
% ~\cite{hess2013:alps,zhang2018, ajello2016}.
(\cite{2021JCAP...02..048A} and references therein).

Here, we focus on the CTA sensitivity to these spectral features using simulated observations of the radio galaxy NGC\,1275, located in the center of the Perseus cluster. We simulate CTA observations of the Perseus cluster and NGC\,1275 to assess its sensitivity to ALP-induced oscillations assuming (i) a 300\,hour exposure during a quiescent state with an intrinsic spectrum equal to the average spectrum observed with MAGIC
% ~\cite{magic-preseus2016} 
and (ii) a 10\,hour exposure during an active state, using the flare spectrum obtained with MAGIC
% ~\cite{2018A&A...617A..91M} 
during an event also followed with VERITAS, as described in \cite{2021JCAP...02..048A}.
% ~\cite{benbow2017}
Generating 100 random realizations of cluster magnetic-field configurations, we numerically calculate, using the \textsc{gammaALPs} code,\footnote{\url{https://github.com/me-manu/gammaALPs}; see also these proceedings} the probability to observe at Earth a $\gamma$ ray of either polarization for an initially unpolarized photon beam.
% The code computes the solution of the equations of motion of the photon-ALP system based on the transfer-matrix formalism and incorporates all relevant terms in the photon-ALP mixing matrix, including the dispersion terms of QED effects and the CMB~\cite{kartavtsev2017}, as well as $\gamma$-ray absorption on the EBL. 

% \subsection{CTA sensitivity to ALP signatures in NGC\,1275 observations}
% \label{sec:alp-results}

\begin{figure}[t]
    \centering
    \includegraphics[width = 0.8\linewidth]{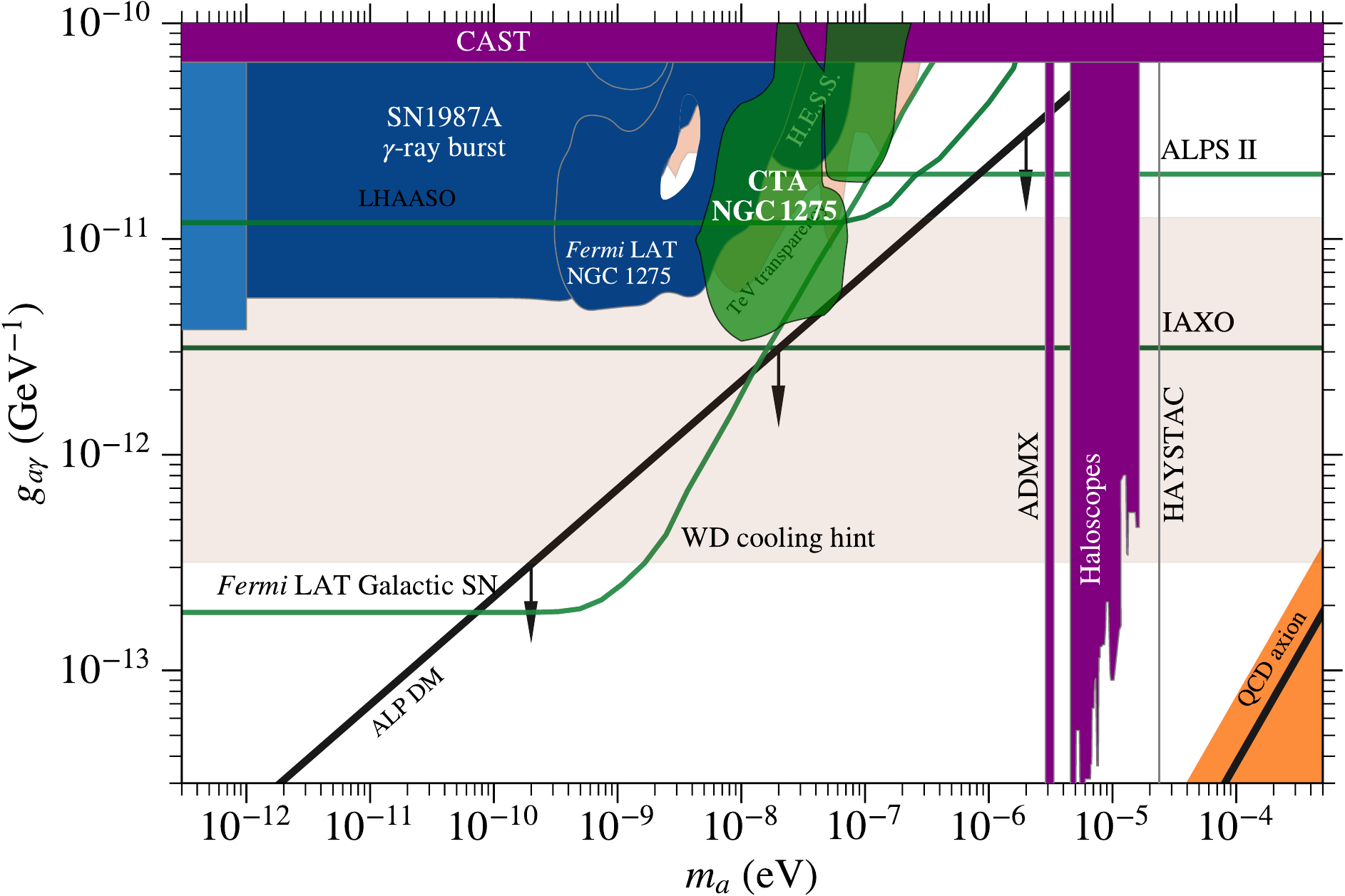}
    \caption{
        \label{fig:alp-summary} Estimated exclusion regions on ALPs from CTA observations (green filled region) compared to other exclusion regions (blue and purple shaded regions) and sensitivities (green lines). 
%         The regions preferred theoretically as well as those where hints for ALPs have been found are also depicted (dark and light orange, respectively). See Ref.~\cite{meyer2016alps} and references therein, updated here with the LHAASO sensitivity derived in Ref.~\cite{vogel2017}. 
%     The figure has been created with the \textsc{gammaALPsPlot} package, see \url{https://github.com/me-manu/gammaALPsPlot}.
        Extracted from~\cite{2021JCAP...02..048A}.
    }
\end{figure}

The resulting upper limits on the detectable gamma ray to ALP coupling $g_{\alpha\gamma}$ obtained for the fiducial cluster magnetic-field setup and the flaring state of NGC\,1275 are compared to other limits and sensitivities in Fig.~\ref{fig:alp-summary}. Notably, CTA observations will improve upon H.E.S.S. limits on photon-ALP coupling by almost an order of magnitude. CTA will also probe ALP masses an order of magnitude higher than those already probed by \emph{Fermi}-LAT observations of the same radio galaxy. Between ${\sim}\,20$\,neV and ${\sim}\,130$\,neV, CTA could deliver the most constraining limits on ALPs to date and will even start exploring the parameter space range where dark matter could consist entirely of ALPs. Furthermore, CTA observations promise to be more sensitive than future searches with LHAASO for the $\gamma$-ray diffuse emission anisotropy above several tens of TeV~\cite{vogel2017}. In the same energy range, CTA observations could also reach a sensitivity similar to future observations with the IAXO and ALPS~II laboratory experiments. Observations of several different sources can be combined to further enhance the CTA sensitivity. At the same time it is worth noting that, in contrast to dedicated laboratory searches, the CTA constraints will be dominated by systematic uncertainties in the model assumptions.

% The mass range probed with CTA is too high to test recent claims for evidence of ALPs~\cite{majumdar2018,xia2018}. This \textbf{evidence} stems from the \textbf{indication} that spectra altered by ALP-induced irregularities  better describe \emph{Fermi}-LAT observations of bright pulsars and Galactic supernova remnants than smooth functions alone. The parameters suggested by these analyses are however incompatible with results from CAST and globular cluster observations~\cite{ayala2014,cast2017}.
% Observations of several different sources can be combined to further enhance the CTA sensitivity. 
% For instance, the AGN IC\,310 in the same Perseus cluster is detected up to 10\,TeV~\cite{magic-ic310}. 
% Through such studies, CTA will become a prime instrument for searching for WISPs dark matter, complementary to searches for more massive candidates.

\section{CTA sensitivity to LIV signatures in blazar spectra}
\label{sec:LIV}

% The precise measurements of blazar spectra with CTA \textbf{also} can serve as a test of Lorentz invariance (\LI). Like any other fundamental principle, exploring its limits of validity has been an important motivation for theoretical and experimental research~\cite{Colladay:1998,Kostelecky:2008}. Lorentz invariance violation (\LIV) is motivated as a possible consequence of theories beyond the Standard Model, such as quantum gravity or string theories (see, {e.g.}, \cite{Alfaro:2005,Kostelecky:1988,Amelino:2001} and Refs.~therein). 

VHE emission and long distances to $\gamma$-ray sources provide a unique opportunity for observational constraints on \LIV signatures. The potential signatures of LIV in the $\gamma$-ray\ band are manifold and include, in particular, energy-dependent time delays, vacuum Cherenkov radiation, photon decay and shifts of the pair-production threshold
(see \cite{2021JCAP...02..048A} and references therein).
% \cite{1998Natur.393..763A,1999ApJ...518L..21K,Stecker:2001,Stecker:2003}. 
In this work, we focus on the CTA potential to test LIV-induced modifications of the pair-production threshold in $\gamma$-ray\ interactions with EBL. If pair-production is affected by LIV, this channel would become a sensitive probe of first- and second-order modifications of dispersion relations. 

Accounting for LIV, the pair production threshold energy for photon-photon head-on collisions is modified as: $\epsilon_{th}'= \frac{m_e^2}{E_{\gamma}'} + \frac{E_{\gamma}'^{n+1}}{4\left(E_{\LIV}^{(n)}\right)^n}$
% \begin{equation}
%     \epsilon_{th}'= \frac{m_e^2}{E_{\gamma}'} + \frac{E_{\gamma}'^{n+1}}{4\left(E_{\LIV}^{(n)}\right)^n}.
% \end{equation}
where $n$ is the LIV leading order. The Lorenz invariance scenario is recovered with $E_{\LIV} \rightarrow \infty$.

In the presence of LIV, the observed source spectrum is modified depending on the value of $E_{\LIV}$ (\cite{2021JCAP...02..048A} and references therein). 
Increased EBL and gamma ray interaction energy thresholds reduce the number of targets that the highest-energy $\gamma$ rays may interact with, increasing the transparency of the Universe to $\gamma$ rays with energies above tens of TeV.

In this work, we investigate the CTA potential to detect LIV on a test case of two blazars, namely Mrk\,501 and 1ES\,0229+200. Both of them have the spectra that may extent beyond several tens of TeV, while being located sufficiently far so that the $\gamma$ rays with energy above few TeV are subject to absorption on EBL. This makes them good targets to search for possible $\gamma$-ray opacity reduction due to LIV presence
% , already proposed as suitable candidates for such searches
% (e.g. \cite{2014JCAP...06..005F, 2016A&A...585A..25T}). 
(e.g. \cite{2014JCAP...06..005F}).
A flaring state of Mrk\,501 is simulated with a 10 hour long CTA exposure, whereas a long-term observation of 1ES\,0229+200 is simulated with a 50 hour one. We assume that the intrinsic spectra of both objects are power laws with exponential cut-offs, whose (comoving) values are set to $E_{\rm cut}' = 10$ and $50$\,TeV correspondingly. These values are compatible with observations of these objects during their extreme states. We use with \gammapy and \ctools to perform these simulations.

To exclude LIV signatures in simulated spectra, we fit the spectra simulated both with and without LIV effects assuming LIV leading orders $n=1,2$. We then compare the best-fit likelihood values to assess the significance of the LIV contribution.
% we compare the best-fit likelihood values for the standard and LIV hypotheses assuming LIV leading orders $n=1,2$. 
We also account for the uncertainty in the EBL intensity in the models used.
% We start by simulating with \gammapy and \ctools spectral observations of CTA affected by LIV in order to determine at which confidence level CTA can detect LIV signatures. We then simulate LI spectra to determine the exclusion capability of CTA. We implemented the LIV effect in the  {\textsc{ebltable}\xspace} Python module\footnote{\url{https://github.com/me- manu/ebltable}} and checked that the difference in best-fit LIV parameters reconstructed with \gammapy and \ctools is small with respect to statistical uncertainties and systematic uncertainties of instrumental or modeling origin. The minor differences are also sub-dominant with respect to the uncertainties induced by the current knowledge of the EBL.
% \subsection{Excluding LIV signal with CTA}
% CTA observations of Mrk\,501 and 1ES\,0229+200 are also simulated assuming that the Lorentz symmetry is not broken. We show the example of the latter \textbf{blazar} in Fig.~\ref{LIV:fig1} (right) for a cut-off at 50\,TeV. We use the profile log-likelihood values for the \LI and \LIV hypotheses to exclude the \LIV scenario at a confidence level given by $\sigma_{\rm rej}^2 = \lambda_{\LIV}-\lambda_{\LI}$. The results for the \LIV leading order $n =1, 2$ are presented in Fig.~\ref{LIV:fig2}, in which we show the capability of CTA to exclude the LIV model as a function of $E_{\LIV}^{(1, 2)}$. The bands correspond to the systematic uncertainties due to the EBL model. 
\begin{figure}[t]
    \centering
    \includegraphics[width=0.49\linewidth]{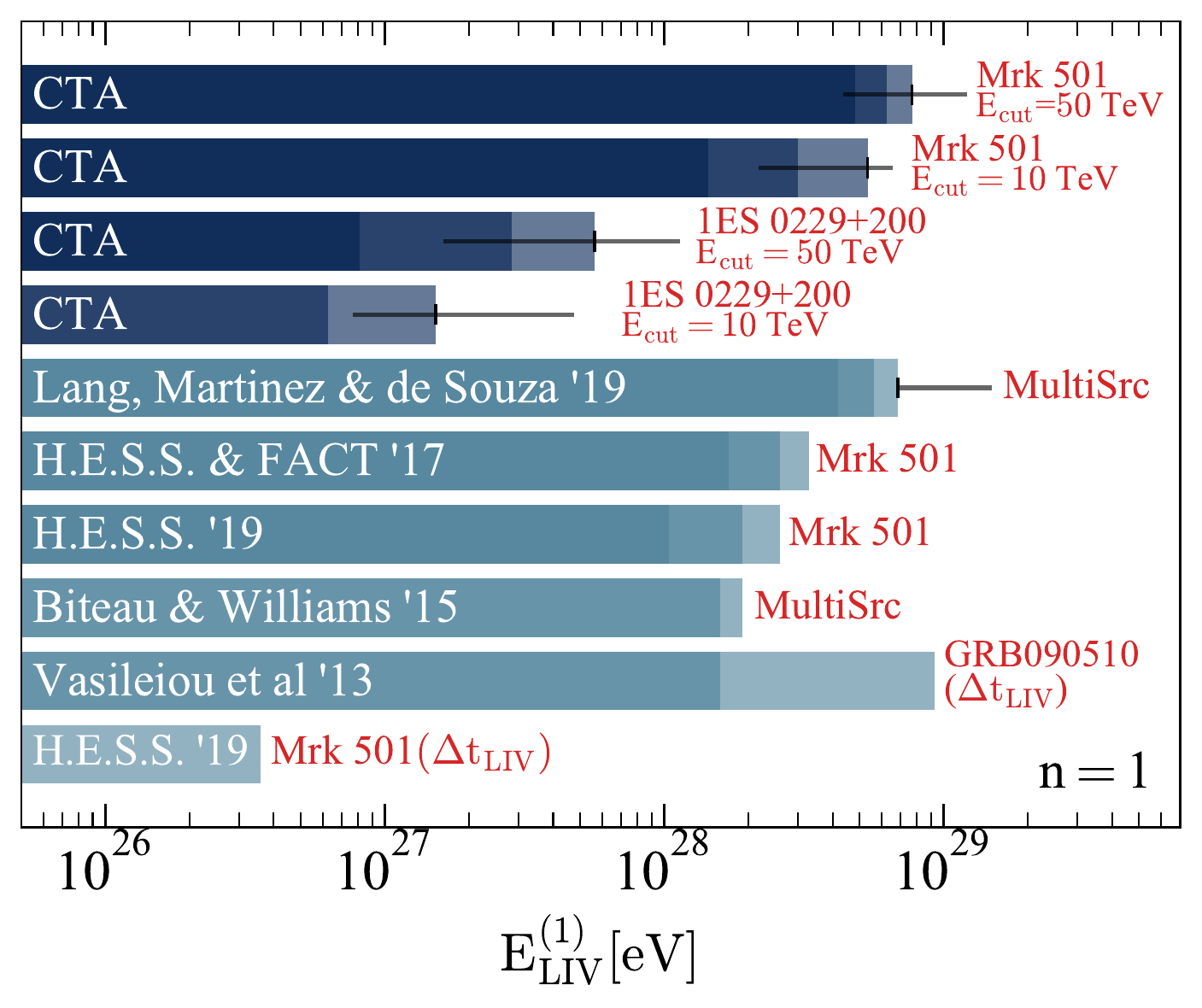}
    \hfill
    \includegraphics[width=0.49\linewidth]{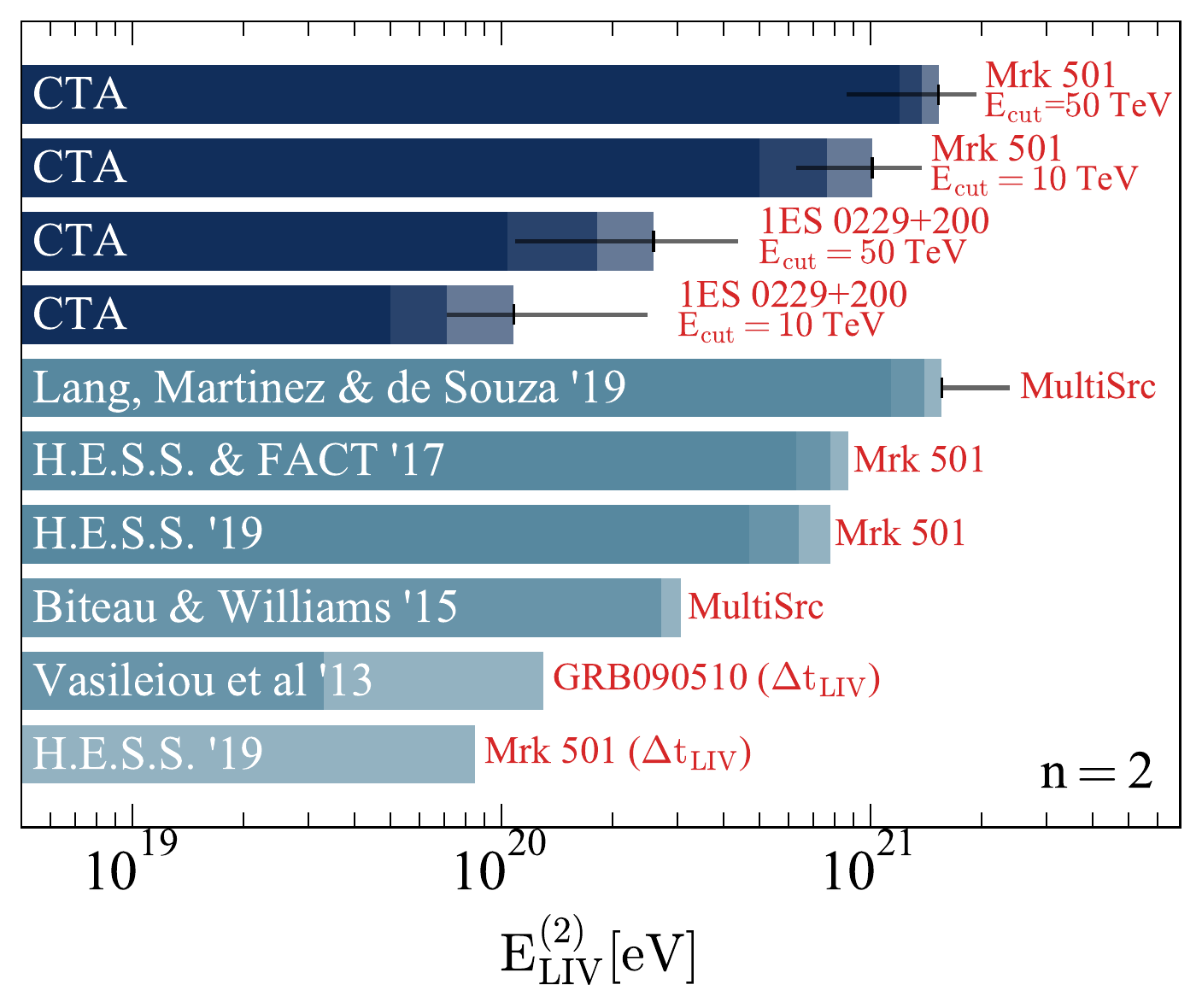}
    \caption{
        \label{LIV:fig4} Comparison of LIV energy scales that can be excluded with CTA and existing subluminal searches in the photon sector. Higher confidence levels in the $2, 3, 5\sigma$ sequence are marked with darker colors. 
%         Limits from Ref.~\cite{2015ApJ...812...60B} (marked as ``Biteau \& Willams '15'') are translated to the photon sector and the quadratic term. Uncertainties on the EBL are already accounted for in the constraints from Ref.~\cite{2015ApJ...812...60B} and shown in black for Ref.~\cite{2019PhRvD..99d3015L} (denoted with ``Lang {\it et al.} '19''). CTA statistical limits from single $\gamma$-ray sources are presented in blue, with black lines marking systematic uncertainties for the 2$\sigma$ limit. Limits on energy-dependent time delays are shown for comparison as annotated with $\Delta t_{\rm LIV}$ \cite{Vasileiou:2013vra,2019ApJ...870...93A}.
        Extracted from~\cite{2021JCAP...02..048A}.
    }
    \label{fig:LIV_lim}
\end{figure}
The resulting LIV energy scales that can be excluded with CTA based on our calculations are shown in Fig.~\ref{fig:LIV_lim}, along with the limits from current-generation instruments using similar analysis techniques.  
% Focusing on constraints from Mrk\,501 during a flaring state (10\,h) and \textbf{from} 1ES\,0229+200 during its quiescent flux state (50\,h), LIV energy scales of $7.7^{+4.4}_{-3.5} \times 10^{28}\,$eV and $5.6^{+5.8}_{-4.0}\times 10^{27}\,$eV could be excluded at the $2\sigma$ confidence level for $n=1$. The quoted uncertainties result from the still-limited constraints on the CIB component of the EBL and from the systematic uncertainties on the \textbf{IRFs}. For $n=2$, the limits are $1.5^{+0.4}_{-0.7} \times 10^{21}\,$eV and $2.6^{+1.8}_{-1.5}\times 10^{20}\,$eV, respectively. 

The predicted CTA limits are more than one order of magnitude better than the recent limits based on energy-dependent time delays stemming from H.E.S.S. observations of Mrk\,501 (``H.E.S.S.'19'' band in Fig.~\ref{fig:LIV_lim}). The CTA limits are also 2-3x more constraining than those obtained by any of currently operating instruments using the same channel and observations of a single source.
Similar prospects can be expected for the multi-source analysis (see the ``Lang~{\it et al.}~'19'' band in the same figure). Complemented with constraints on time delays from blazars, $\gamma$-ray bursts and pulsars, this makes CTA a promising explorer of fundamental symmetries in the photon sector at energy scales that remain beyond the reach of accelerator-based experiments.

\bibliographystyle{jhep}
\bibliography{references}

% Full authors list (ONLY FOR COLLABORATIONS)
% \clearpage
% \section*{Full Authors List: \Coll\ Collaboration}

%\noindent \textbf{Note comment afterwards:} Collaborations have the possibility to provide an authors list in xml format which will be used while generating the DOI entries making the full authors list searchable in databases like Inspire HEP. For instructions please go to icrc2021.desy.de/proceedings or contact us under icrc2021proc@desy.de.\\
%
%\scriptsize
%\noindent
%first.author$^1$, 
%second.author$^2$, 
%third.author$^3$ % .... more names
%and 
%last.author$^{n}$ \\
%
%\noindent
%$^1$first.affiliation.
%$^2$second.affiliation. % .... more affiliation
%$^{m}$last.affiliation.

%\title{A CTA Consortium paper

\clearpage
\section*{Full Authors List: \Coll\ Collaboration}

\scriptsize
\noindent
H.~Abdalla,$^{1}$
H.~Abe,$^{2}$
F.~Acero,$^{3}$
A.~Acharyya,$^{4}$
R.~Adam,$^{5}$
I.~Agudo,$^{6}$
A.~Aguirre-Santaella,$^{7}$
R.~Alfaro,$^{8}$
J.~Alfaro,$^{9}$
C.~Alispach,$^{10}$
% ...

\bigskip 
\bigskip
\noindent
\noindent $^{1} \ $Centre for Space Research, North-West University, Potchefstroom, 2520, South Africa

\noindent $^{2} \ $Institute for Cosmic Ray Research, University of Tokyo, 5-1-5, Kashiwa-no-ha, Kashiwa, Chiba 277-8582, Japan

\noindent $^{3} \ $AIM, CEA, CNRS, Université Paris-Saclay, Université Paris Diderot, Sorbonne Paris Cité, CEA Paris-Saclay, IRFU/DAp, Bat 709, Orme des Merisiers, 91191 Gif-sur-Yvette, France

\noindent $^{4} \ $Centre for Advanced Instrumentation, Dept. of Physics, Durham University, South Road, Durham DH1 3LE, United Kingdom

\noindent $^{5} \ $Laboratoire Leprince-Ringuet, École Polytechnique (UMR 7638, CNRS/IN2P3, Institut Polytechnique de Paris), 91128 Palaiseau, France

\noindent $^{6} \ $Instituto de Astrofísica de Andalucía-CSIC, Glorieta de la Astronomía s/n, 18008, Granada, Spain

\noindent $^{7} \ $Instituto de Física Teórica UAM/CSIC and Departamento de Física Teórica, Universidad Autónoma de Madrid, c/ Nicolás Cabrera 13-15, Campus de Cantoblanco UAM, 28049 Madrid, Spain

\noindent $^{8} \ $Universidad Nacional Autónoma de México, Delegación Coyoacán, 04510 Ciudad de México, Mexico

\noindent $^{9} \ $Pontificia Universidad Católica de Chile, Av. Libertador Bernardo O'Higgins 340, Santiago, Chile

\noindent $^{10} \ $University of Geneva - Département de physique nucléaire et corpusculaire, 24 rue du Général-Dufour, 1211 Genève 4, Switzerland

\noindent $^{11} \ $INFN Dipartimento di Scienze Fisiche e Chimiche - Università degli Studi dell'Aquila and Gran Sasso Science Institute, Via Vetoio 1, Viale Crispi 7, 67100 L'Aquila, Italy

\noindent $^{12} \ $Instituto de Astronomia, Geofísico, e Ciências Atmosféricas - Universidade de São Paulo, Cidade Universitária, R. do Matão, 1226, CEP 05508-090, São Paulo, SP, Brazil

\noindent $^{13} \ $INAF - Osservatorio di Astrofisica e Scienza dello spazio di Bologna, Via Piero Gobetti 93/3, 40129  Bologna, Italy

\noindent $^{14} \ $INAF - Osservatorio Astrofisico di Arcetri, Largo E. Fermi, 5 - 50125 Firenze, Italy

\noindent $^{15} \ $INFN Sezione di Perugia and Università degli Studi di Perugia, Via A. Pascoli, 06123 Perugia, Italy

\noindent $^{16} \ $Aix Marseille Univ, CNRS/IN2P3, CPPM, 163 Avenue de Luminy, 13288 Marseille cedex 09, France

\noindent $^{17} \ $FZU - Institute of Physics of the Czech Academy of Sciences, Na Slovance 1999/2, 182 21 Praha 8, Czech Republic

\noindent $^{18} \ $Astronomical Institute of the Czech Academy of Sciences, Bocni II 1401 - 14100 Prague, Czech Republic

\noindent $^{19} \ $EMFTEL department  and IPARCOS, Universidad Complutense de Madrid, 28040 Madrid, Spain

\noindent $^{20} \ $Laboratoire Univers et Particules de Montpellier, Université de Montpellier, CNRS/IN2P3, CC 72, Place Eugène Bataillon, F-34095 Montpellier Cedex 5, France

\noindent $^{21} \ $School of Physics, University of New South Wales, Sydney NSW 2052, Australia

\noindent $^{22} \ $University of Namibia, Department of Physics, 340 Mandume Ndemufayo Ave., Pioneerspark, Windhoek, Namibia

\noindent $^{23} \ $School of Physics and Astronomy, Monash University, Melbourne, Victoria 3800, Australia

\noindent $^{24} \ $Department of Astronomy, University of Geneva, Chemin d'Ecogia 16, CH-1290 Versoix, Switzerland

\noindent $^{25} \ $INAF - Osservatorio Astrofisico di Torino, Strada Osservatorio 20, 10025  Pino Torinese (TO), Italy

\noindent $^{26} \ $Deutsches Elektronen-Synchrotron, Platanenallee 6, 15738 Zeuthen, Germany

\noindent $^{27} \ $RIKEN, Institute of Physical and Chemical Research, 2-1 Hirosawa, Wako, Saitama, 351-0198, Japan

\noindent $^{28} \ $Centro Brasileiro de Pesquisas Físicas, Rua Xavier Sigaud 150, RJ 22290-180, Rio de Janeiro, Brazil

\noindent $^{29} \ $Instituto de Astrofísica de Canarias and Departamento de Astrofísica, Universidad de La Laguna, La Laguna, Tenerife, Spain

\noindent $^{30} \ $Department of Physics and Electrical Engineering, Linnaeus University, 351 95 Växjö, Sweden

\noindent $^{31} \ $University of the Witwatersrand, 1 Jan Smuts Avenue, Braamfontein, 2000 Johannesburg, South Africa

\noindent $^{32} \ $Institut für Theoretische Physik, Lehrstuhl IV: Plasma-Astroteilchenphysik, Ruhr-Universität Bochum, Universitätsstraße 150, 44801 Bochum, Germany

\noindent $^{33} \ $Center for Astrophysics | Harvard \& Smithsonian, 60 Garden St, Cambridge, MA 02180, USA

\noindent $^{34} \ $INFN Sezione di Torino, Via P. Giuria 1, 10125 Torino, Italy

\noindent $^{35} \ $Finnish Centre for Astronomy with ESO, University of Turku, Finland, FI-20014 University of Turku, Finland

\noindent $^{36} \ $Pidstryhach Institute for Applied Problems in Mechanics and Mathematics NASU, 3B Naukova Street, Lviv, 79060, Ukraine

\noindent $^{37} \ $Institut für Astronomie und Astrophysik, Universität Tübingen, Sand 1, 72076 Tübingen, Germany

\noindent $^{38} \ $Laboratoire de Physique des 2 infinis, Irene Joliot-Curie, IN2P3/CNRS, Université Paris-Saclay, Université de Paris, 15 rue Georges Clemenceau, 91406 Orsay, Cedex, France

\noindent $^{39} \ $ETH Zurich, Institute for Particle Physics, Schafmattstr. 20, CH-8093 Zurich, Switzerland

\noindent $^{40} \ $INFN Sezione di Bari and Politecnico di Bari, via Orabona 4, 70124 Bari, Italy

\noindent $^{41} \ $Institut de Fisica d'Altes Energies (IFAE), The Barcelona Institute of Science and Technology, Campus UAB, 08193 Bellaterra (Barcelona), Spain

\noindent $^{42} \ $INAF - Osservatorio Astronomico di Palermo "G.S. Vaiana", Piazza del Parlamento 1, 90134 Palermo, Italy

\noindent $^{43} \ $LUTH, GEPI and LERMA, Observatoire de Paris, CNRS, PSL University, 5 place Jules Janssen, 92190, Meudon, France

\noindent $^{44} \ $Sorbonne Université, Université Paris Diderot, Sorbonne Paris Cité, CNRS/IN2P3, Laboratoire de Physique Nucléaire et de Hautes Energies, LPNHE, 4 Place Jussieu, F-75005 Paris, France

\noindent $^{45} \ $INAF - Osservatorio Astrofisico di Catania, Via S. Sofia, 78, 95123 Catania, Italy

\noindent $^{46} \ $Instituto de Física de São Carlos, Universidade de São Paulo, Av. Trabalhador São-carlense, 400 - CEP 13566-590, São Carlos, SP, Brazil

\noindent $^{47} \ $INFN and Università degli Studi di Siena, Dipartimento di Scienze Fisiche, della Terra e dell'Ambiente (DSFTA), Sezione di Fisica, Via Roma 56, 53100 Siena, Italy

\noindent $^{48} \ $Departament de Física Quàntica i Astrofísica, Institut de Ciències del Cosmos, Universitat de Barcelona, IEEC-UB, Martí i Franquès, 1, 08028, Barcelona, Spain

\noindent $^{49} \ $INFN Sezione di Padova and Università degli Studi di Padova, Via Marzolo 8, 35131 Padova, Italy

\noindent $^{50} \ $Astronomy Department of Faculty of Physics, Sofia University, 5 James Bourchier Str., 1164 Sofia, Bulgaria

\noindent $^{51} \ $Department of Physics, Columbia University, 538 West 120th Street, New York, NY 10027, USA

\noindent $^{52} \ $INAF - Istituto di Astrofisica Spaziale e Fisica Cosmica di Palermo, Via U. La Malfa 153, 90146 Palermo, Italy

\noindent $^{53} \ $Armagh Observatory and Planetarium, College Hill, Armagh BT61 9DG, United Kingdom

\noindent $^{54} \ $INFN Sezione di Catania, Via S. Sofia 64, 95123 Catania, Italy

\noindent $^{55} \ $INAF - Osservatorio Astronomico di Brera, Via Brera 28, 20121 Milano, Italy

\noindent $^{56} \ $Kavli Institute for Particle Astrophysics and Cosmology, Department of Physics and SLAC National Accelerator Laboratory, Stanford University, 2575 Sand Hill Road, Menlo Park, CA 94025, USA

\noindent $^{57} \ $Universidade Cruzeiro do Sul, Núcleo de Astrofísica Teórica (NAT/UCS), Rua Galvão Bueno 8687, Bloco B, sala 16, Libertade 01506-000 - São Paulo, Brazil

\noindent $^{58} \ $INAF - Osservatorio Astronomico di Roma, Via di Frascati 33, 00040, Monteporzio Catone, Italy

\noindent $^{59} \ $INAF - Istituto di Astrofisica Spaziale e Fisica Cosmica di Milano, Via A. Corti 12, 20133 Milano, Italy

\noindent $^{60} \ $INFN Sezione di Pisa, Largo Pontecorvo 3, 56217 Pisa, Italy

\noindent $^{61} \ $The Henryk Niewodniczański Institute of Nuclear Physics, Polish Academy of Sciences, ul. Radzikowskiego 152, 31-342 Cracow, Poland

\noindent $^{62} \ $Max-Planck-Institut für Kernphysik, Saupfercheckweg 1, 69117 Heidelberg, Germany

\noindent $^{63} \ $INAF - Osservatorio Astronomico di Capodimonte, Via Salita Moiariello 16, 80131 Napoli, Italy

\noindent $^{64} \ $INFN Sezione di Trieste and Università degli Studi di Udine, Via delle Scienze 208, 33100 Udine, Italy

\noindent $^{65} \ $Palacky University Olomouc, Faculty of Science, RCPTM, 17. listopadu 1192/12, 771 46 Olomouc, Czech Republic

\noindent $^{66} \ $Dublin City University, Glasnevin, Dublin 9, Ireland

\noindent $^{67} \ $Dipartimento di Fisica - Universitá degli Studi di Torino, Via Pietro Giuria 1 - 10125 Torino, Italy

\noindent $^{68} \ $CIEMAT, Avda. Complutense 40, 28040 Madrid, Spain

\noindent $^{69} \ $Universidad de Valparaíso, Blanco 951, Valparaiso, Chile

\noindent $^{70} \ $INAF - Istituto di Radioastronomia, Via Gobetti 101, 40129 Bologna, Italy

\noindent $^{71} \ $University of Oxford, Department of Physics, Denys Wilkinson Building, Keble Road, Oxford OX1 3RH, United Kingdom

\noindent $^{72} \ $Cherenkov Telescope Array Observatory, Saupfercheckweg 1, 69117 Heidelberg, Germany

\noindent $^{73} \ $LAPP, Univ. Grenoble Alpes, Univ. Savoie Mont Blanc, CNRS-IN2P3, 9 Chemin de Bellevue - BP 110, 74941 Annecy Cedex, France

\noindent $^{74} \ $Universidade Federal Do Paraná - Setor Palotina, Departamento de Engenharias e Exatas, Rua Pioneiro, 2153, Jardim Dallas, CEP: 85950-000 Palotina, Paraná, Brazil

\noindent $^{75} \ $Institute of Space Sciences (ICE-CSIC), and Institut d'Estudis Espacials de Catalunya (IEEC), and Institució Catalana de Recerca I Estudis Avançats (ICREA), Campus UAB, Carrer de Can Magrans, s/n 08193 Cerdanyola del Vallés, Spain

\noindent $^{76} \ $INFN Sezione di Napoli, Via Cintia, ed. G, 80126 Napoli, Italy

\noindent $^{77} \ $Universitá degli Studi di Napoli "Federico II" - Dipartimento di Fisica "E. Pancini", Complesso universitario di Monte Sant'Angelo, Via Cintia - 80126 Napoli, Italy

\noindent $^{78} \ $Université de Paris, CNRS, Astroparticule et Cosmologie, 10, rue Alice Domon et Léonie Duquet, 75013 Paris Cedex 13, France

\noindent $^{79} \ $Institute for Theoretical Physics and Astrophysics, Universität Würzburg, Campus Hubland Nord, Emil-Fischer-Str. 31, 97074 Würzburg, Germany

\noindent $^{80} \ $Enrico Fermi Institute, University of Chicago, 5640 South Ellis Avenue, Chicago, IL 60637, USA

\noindent $^{81} \ $Center for Astrophysics and Cosmology, University of Nova Gorica, Vipavska 11c, 5270 Ajdovščina, Slovenia

\noindent $^{82} \ $School of Physical Sciences, University of Adelaide, Adelaide SA 5005, Australia

\noindent $^{83} \ $Department of Physics, TU Dortmund University, Otto-Hahn-Str. 4, 44221 Dortmund, Germany

\noindent $^{84} \ $King's College London, Strand, London, WC2R 2LS, United Kingdom

\noindent $^{85} \ $Escola de Artes, Ciências e Humanidades, Universidade de São Paulo, Rua Arlindo Bettio, CEP 03828-000, 1000 São Paulo, Brazil

\noindent $^{86} \ $Unitat de Física de les Radiacions, Departament de Física, and CERES-IEEC, Universitat Autònoma de Barcelona, Edifici C3, Campus UAB, 08193 Bellaterra, Spain

\noindent $^{87} \ $Grupo de Electronica, Universidad Complutense de Madrid, Av. Complutense s/n, 28040 Madrid, Spain

\noindent $^{88} \ $Department of Physical Science, Hiroshima University, Higashi-Hiroshima, Hiroshima 739-8526, Japan

\noindent $^{89} \ $Department of Physics, Nagoya University, Chikusa-ku, Nagoya, 464-8602, Japan

\noindent $^{90} \ $Alikhanyan National Science Laboratory, Yerevan Physics Institute, 2 Alikhanyan Brothers St., 0036, Yerevan, Armenia

\noindent $^{91} \ $INFN Sezione di Bari and Università degli Studi di Bari, via Orabona 4, 70124 Bari, Italy

\noindent $^{92} \ $IRFU, CEA, Université Paris-Saclay, Bât 141, 91191 Gif-sur-Yvette, France

\noindent $^{93} \ $Universidad Andres Bello, República 252, Santiago, Chile

\noindent $^{94} \ $Academic Computer Centre CYFRONET AGH, ul. Nawojki 11, 30-950 Cracow, Poland

\noindent $^{95} \ $Department of Natural Sciences, The Open University of Israel, 1 University Road, POB 808, Raanana 43537, Israel

\noindent $^{96} \ $Astronomy Department, Adler Planetarium and Astronomy Museum, Chicago, IL 60605, USA

\noindent $^{97} \ $Department of Physics, Yamagata University, Yamagata, Yamagata 990-8560, Japan

\noindent $^{98} \ $Tohoku University, Astronomical Institute, Aobaku, Sendai 980-8578, Japan

\noindent $^{99} \ $Santa Cruz Institute for Particle Physics and Department of Physics, University of California, Santa Cruz, 1156 High Street, Santa Cruz, CA 95064, USA

\noindent $^{100} \ $Astronomical Observatory of Taras Shevchenko National University of Kyiv, 3 Observatorna Street, Kyiv, 04053, Ukraine

\noindent $^{101} \ $Department of Physics and Astronomy and the Bartol Research Institute, University of Delaware, Newark, DE 19716, USA

\noindent $^{102} \ $IMAPP, Radboud University Nijmegen, P.O. Box 9010, 6500 GL Nijmegen, The Netherlands

\noindent $^{103} \ $Josip Juraj Strossmayer University of Osijek, Trg Ljudevita Gaja 6, 31000 Osijek, Croatia

\noindent $^{104} \ $Max-Planck-Institut für Physik, Föhringer Ring 6, 80805 München, Germany

\noindent $^{105} \ $INFN Sezione di Bari, via Orabona 4, 70126 Bari, Italy

\noindent $^{106} \ $INFN Sezione di Roma La Sapienza, P.le Aldo Moro, 2 - 00185 Roma, Italy

\noindent $^{107} \ $Astronomical Observatory, Jagiellonian University, ul. Orla 171, 30-244 Cracow, Poland

\noindent $^{108} \ $University of Alabama, Tuscaloosa, Department of Physics and Astronomy, Gallalee Hall, Box 870324 Tuscaloosa, AL 35487-0324, USA

\noindent $^{109} \ $Friedrich-Alexander-Universit\"at Erlangen-N\"urnberg, Erlangen Centre for Astroparticle Physics (ECAP), Erwin-Rommel-Str. 1, 91058 Erlangen, Germany

\noindent $^{110} \ $University of Iowa, Department of Physics and Astronomy, Van Allen Hall, Iowa City, IA 52242, USA

\noindent $^{111} \ $Division of Physics and Astronomy, Graduate School of Science, Kyoto University, Sakyo-ku, Kyoto, 606-8502, Japan

\noindent $^{112} \ $Institut für Astro- und Teilchenphysik, Leopold-Franzens-Universität, Technikerstr. 25/8, 6020 Innsbruck, Austria

\noindent $^{113} \ $Institut de Recherche en Astrophysique et Planétologie, CNRS-INSU, Université Paul Sabatier, 9 avenue Colonel Roche, BP 44346, 31028 Toulouse Cedex 4, France

\noindent $^{114} \ $Institute of Particle and Nuclear Studies,  KEK (High Energy Accelerator Research Organization), 1-1 Oho, Tsukuba, 305-0801, Japan

\noindent $^{115} \ $Dept. of Physics and Astronomy, University of Leicester, Leicester, LE1 7RH, United Kingdom

\noindent $^{116} \ $CENBG, Univ. Bordeaux, CNRS-IN2P3, UMR 5797, 19 Chemin du Solarium, CS 10120, F-33175 Gradignan Cedex, France

\noindent $^{117} \ $Dipartimento di Fisica e Astronomia, Sezione Astrofisica, Universitá di Catania, Via S. Sofia 78, I-95123 Catania, Italy

\noindent $^{118} \ $Department of Physics, Humboldt University Berlin, Newtonstr. 15, 12489 Berlin, Germany

\noindent $^{119} \ $INFN Sezione di Trieste and Università degli Studi di Trieste, Via Valerio 2 I, 34127 Trieste, Italy

\noindent $^{120} \ $National Centre for nuclear research (Narodowe Centrum Badań Jądrowych), Ul. Andrzeja Sołtana7, 05-400 Otwock, Świerk, Poland

\noindent $^{121} \ $Institute for Nuclear Research and Nuclear Energy, Bulgarian Academy of Sciences, 72 boul. Tsarigradsko chaussee, 1784 Sofia, Bulgaria

\noindent $^{122} \ $University of Rijeka, Department of Physics, Radmile Matejcic 2,  51000 Rijeka, Croatia

\noindent $^{123} \ $University of Białystok, Faculty of Physics, ul. K. Ciołkowskiego 1L, 15-254 Białystok, Poland

\noindent $^{124} \ $Anton Pannekoek Institute/GRAPPA, University of Amsterdam, Science Park 904 1098 XH Amsterdam, The Netherlands

\noindent $^{125} \ $Escuela Politécnica Superior de Jaén, Universidad de Jaén, Campus Las Lagunillas s/n, Edif. A3, 23071 Jaén, Spain

\noindent $^{126} \ $Physik-Institut, Universität Zürich, Winterthurerstrasse 190, 8057 Zürich, Switzerland

\noindent $^{127} \ $Hiroshima Astrophysical Science Center, Hiroshima University, Higashi-Hiroshima, Hiroshima 739-8526, Japan

\noindent $^{128} \ $University of Wisconsin, Madison, 500 Lincoln Drive, Madison, WI, 53706, USA

\noindent $^{129} \ $Nicolaus Copernicus Astronomical Center, Polish Academy of Sciences, ul. Bartycka 18, 00-716 Warsaw, Poland

\noindent $^{130} \ $INFN Sezione di Roma Tor Vergata, Via della Ricerca Scientifica 1, 00133 Rome, Italy

\noindent $^{131} \ $Department of Physics, University of Bath, Claverton Down, Bath BA2 7AY, United Kingdom

\noindent $^{132} \ $School of Allied Health Sciences, Kitasato University, Sagamihara, Kanagawa 228-8555, Japan

\noindent $^{133} \ $Department of Physics, Tokai University, 4-1-1, Kita-Kaname, Hiratsuka, Kanagawa 259-1292, Japan

\noindent $^{134} \ $Charles University, Institute of Particle \& Nuclear Physics, V Holešovičkách 2, 180 00 Prague 8, Czech Republic

\noindent $^{135} \ $Graduate School of Science, University of Tokyo, 7-3-1 Hongo, Bunkyo-ku, Tokyo 113-0033, Japan

\noindent $^{136} \ $Department of Physics and Astronomy, University of California, Los Angeles, CA 90095, USA

\noindent $^{137} \ $Graduate School of Technology, Industrial and Social Sciences, Tokushima University, Tokushima 770-8506, Japan

\noindent $^{138} \ $Instituto de Física - Universidade de São Paulo, Rua do Matão Travessa R Nr.187 CEP 05508-090  Cidade Universitária, São Paulo, Brazil

\noindent $^{139} \ $Institut für Physik \& Astronomie, Universität Potsdam, Karl-Liebknecht-Strasse 24/25, 14476 Potsdam, Germany

\noindent $^{140} \ $International Institute of Physics at the Federal University of Rio Grande do Norte, Campus Universitário, Lagoa Nova CEP 59078-970 Rio Grande do Norte, Brazil

\noindent $^{141} \ $Landessternwarte, Zentrum für Astronomie  der Universität Heidelberg, Königstuhl 12, 69117 Heidelberg, Germany

\noindent $^{142} \ $University of Johannesburg, Department of Physics, University Road, PO Box 524, Auckland Park 2006, South Africa

\noindent $^{143} \ $Departamento de Astronomía, Universidad de Concepción, Barrio Universitario S/N, Concepción, Chile

\noindent $^{144} \ $National Astronomical Research Institute of Thailand, 191 Huay Kaew Rd., Suthep, Muang, Chiang Mai, 50200, Thailand

\noindent $^{145} \ $Space Research Centre, Polish Academy of Sciences, ul. Bartycka 18A, 00-716 Warsaw, Poland

\noindent $^{146} \ $The University of Manitoba, Dept of Physics and Astronomy, Winnipeg, Manitoba R3T 2N2, Canada

\noindent $^{147} \ $Institute of Astronomy, Faculty of Physics, Astronomy and Informatics, Nicolaus Copernicus University in Toruń, ul. Grudziądzka 5, 87-100 Toruń, Poland

\noindent $^{148} \ $University of Oslo, Department of Physics, Sem Saelandsvei 24 - PO Box 1048 Blindern, N-0316 Oslo, Norway

\noindent $^{149} \ $Western Sydney University, Locked Bag 1797, Penrith, NSW 2751, Australia

\noindent $^{150} \ $Agenzia Spaziale Italiana (ASI), 00133 Roma, Italy

\noindent $^{151} \ $University of Hawai'i at Manoa, 2500 Campus Rd, Honolulu, HI, 96822, USA

\noindent $^{152} \ $University of Groningen, KVI - Center for Advanced Radiation Technology, Zernikelaan 25, 9747 AA Groningen, The Netherlands

\noindent $^{153} \ $Department of Physics and Mathematics, Aoyama Gakuin University, Fuchinobe, Sagamihara, Kanagawa, 252-5258, Japan

\noindent $^{154} \ $Faculty of Science, Ibaraki University, Mito, Ibaraki, 310-8512, Japan

\noindent $^{155} \ $University of Split  - FESB, R. Boskovica 32, 21 000 Split, Croatia

\noindent $^{156} \ $School of Physics \& Center for Relativistic Astrophysics, Georgia Institute of Technology, 837 State Street, Atlanta, Georgia, 30332-0430, USA

\noindent $^{157} \ $Escuela Politécnica Superior de Jaén, Universidad de Jaén, Campus Las Lagunillas s/n, Edif. A3, 23071 Jaén, Spain

\end{document}